\documentclass[dvips]{article}

\usepackage{icrc2011}

\title{The forward particle production in the energy range of 1 PeV as seen with the Tibet hybrid experiment}

\newcommand{\etal}{\MakeLowercase{\textit{et al. }}} 
\shorttitle{Ying~Zhang \etal The forward particle production in the
energy range of $10^{15}$eV}

\authors{M.~Amenomori$^{1}$, X.~J.~Bi$^{2}$, D.~Chen$^{3}$, W.~Y.~Chen$^{2}$, S.~W.~Cui$^{4}$,
Danzengluobu$^{5}$, L.~K.~Ding$^{2}$, X.~H.~Ding$^{5}$,
C.~F.~Feng$^{6}$, Zhaoyang Feng$^{2}$, Z.~Y.~Feng$^{7}$,
Q.~B.~Gou$^{2}$, H.~W.~Guo$^{5}$, Y.~Q.~Guo$^{2}$, H.~H.~He$^{2}$,
Z.~T.~He$^{4,2}$, K.~Hibino$^{8}$, N.~Hotta$^{9}$, Haibing~Hu$^{5}$,
H.~B.~Hu$^{2}$, J.~Huang$^{2}$, W.~J.~Li$^{2,7}$, H.~Y.~Jia$^{7}$,
L.~Jiang$^{2}$, F.~Kajino$^{10}$, K.~Kasahara$^{11}$,
Y.~Katayose$^{12}$, C.~Kato$^{13}$, K.~Kawata$^{3}$,
Labaciren$^{5}$, G.~M.~Le$^{2}$, A.~F.~Li$^{14,6,2}$, C.~Liu$^{2}$,
J.~S.~Liu$^{2}$, H.~Lu$^{2}$, X.~R.~Meng$^{5}$,
K.~Mizutani$^{11,15}$, K.~Munakata$^{13}$, H.~Nanjo$^{1}$,
M.~Nishizawa$^{16}$, M.~Ohnishi$^{3}$, I.~Ohta$^{17}$,
S.~Ozawa$^{11}$, X.~L.~Qian$^{6,2}$, X.~B.~Qu$^{2}$,
T.~Saito$^{18}$, T.~Y.~Saito$^{19}$, M.~Sakata$^{10}$,
T.~K.~Sako$^{12}$, J.~Shao$^{2,6}$, M.~Shibata$^{12}$,
A.~Shiomi$^{20}$, T.~Shirai$^{8}$, H.~Sugimoto$^{21}$,
M.~Takita$^{3}$, Y.~H.~Tan$^{2}$, N.~Tateyama$^{8}$,
S.~Torii$^{11}$, H.~Tsuchiya$^{22}$, S.~Udo$^{8}$, H.~Wang$^{2}$,
H.~R.~Wu$^{2}$, L.~Xue$^{6}$, Y.~Yamamoto$^{10}$, Z.~Yang$^{2}$,
S.~Yasue$^{23}$, A.~F.~Yuan$^{5}$, T.~Yuda$^{3}$, L.~M.~Zhai$^{2}$,
H.~M.~Zhang$^{2}$, J.~L.~Zhang$^{2}$, X.~Y.~Zhang$^{6}$,
Y.~Zhang$^{2}$, Yi~Zhang$^{2}$, Ying~Zhang$^{2}$,
Zhaxisangzhu$^{5}$, X.~X.~Zhou$^{7}$\ (The Tibet AS$\gamma$
Collaboration)}
\afiliations{$^{1}$Department of Physics, Hirosaki University, Hirosaki 036-8561, Japan\\
$^{2}$Key Laboratory of Particle Astrophysics, Institute of High
Energy Physics, Chinese
Academy of Sciences, Beijing 100049, China\\
$^{3}$Institute for Cosmic Ray Research, University of Tokyo, Kashiwa 277-8582, Japan\\
$^{4}$Department of Physics, Hebei Normal University, Shijiazhuang 050016, China\\
$^{5}$Department of Mathematics and Physics, Tibet University, Lhasa 850000, China\\
$^{6}$Department of Physics, Shandong University, Jinan 250100, China\\
$^{7}$Institute of Modern Physics, SouthWest Jiaotong University, Chengdu 610031, China\\
$^{8}$Faculty of Engineering, Kanagawa University, Yokohama 221-8686, Japan\\
$^{9}$Faculty of Education, Utsunomiya University, Utsunomiya 321-8505, Japan\\
$^{10}$Department of Physics, Konan University, Kobe 658-8501, Japan\\
$^{11}$Research Institute for Science and Engineering, Waseda
University, Tokyo 169-8555,
Japan\\
$^{12}$Faculty of Engineering, Yokohama National University, Yokohama 240-8501, Japan\\
$^{13}$Department of Physics, Shinshu University, Matsumoto 390-8621, Japan\\
$^{14}$School of Information Science and Engineering, Shandong
Agriculture University,
Taian 271018, China\\
$^{15}$Saitama University, Saitama 338-8570, Japan\\
$^{16}$National Institute of Informatics, Tokyo 101-8430, Japan\\
$^{17}$Sakushin Gakuin University, Utsunomiya 321-3295, Japan\\
$^{18}$Tokyo Metropolitan College of Industrial Technology, Tokyo 116-8523, Japan\\
$^{19}$Max-Planck-Institut f\"ur Physik, M\"unchen D-80805, Deutschland\\
$^{20}$College of Industrial Technology, Nihon University, Narashino 275-8576, Japan\\
$^{21}$Shonan Institute of Technology, Fujisawa 251-8511, Japan\\
$^{22}$RIKEN, Wako 351-0198, Japan\\
$^{23}$School of General Education, Shinshu University, Matsumoto
390-8621, Japan}

\abstract{We are now operating the 500 $m^2$ Yangbajing air-shower
core (YAC-II) array near the center of the Tibet air-shower array (
Tibet-III ) to observe cosmic-ray chemical composition at the knee
energy region since February 2011. The first step of YAC, called
¡°YAC-I¡±, containing 16 detector units, was operated from May, 2009
to February, 2010. In this paper, we used the YAC-I and Tibet-III
coincident data set obtained from May, 2009 through January, 2010 to
present the electromagnetic spectrum of air shower cores at around
$10^{15}$ eV energy region. The effective live time is calculated as
100.5 days. We would like to report the comparison of our
experimental data with MC model prediction in this paper.}

\begin{document}
\maketitle

\section{Introduction}

$\;\;\;\;$Direct measurements of the primary cosmic rays (CR) with
energies higher than $10^{15}$ eV are difficult due to their low
flux and the limited detector acceptance of the on board satellite
or balloon experiment. Instead, their properties are reconstructed
from the measurements of the extensive air showers (EAS) they
produce in the atmosphere. The reconstruction of EAS events is based
on Monte Carlo hadronic interaction models of the air shower
development, which are based on the knowledge obtained from the
accelerator hadron-nucleus collision experiments. Since accelerator
experiment can not provide all information that cosmic ray studies
need, some extrapolation to higher energies and to un-reached phase
space is inevitable that induces uncertainty in the explanation of
AS phenomenon.

$\;\;\;\;$It is well known that the produced particles in the most
forward region of hadronic interactions are most responsible for the
AS development, and the most forward region is the dead-corner of
conventional collider experiments. It is also known that the high
energy particles in the AS core region are most sensitive to the
forward region particle productions. Because of the advantage
observing EAS cores in the high altitude, a new hybrid experiment
was constructed and operated in Yangbajing, Tibet.

$\;\;\;\;$In present work, we report the checking of hadronic
interaction models by observing EAS cores at the energy region of
$10^{15}$ eV using Yangbajing Air shower Core detectors (YAC-I) and
the air-shower array (Tibet-III).

\vspace{-0.3cm}
\section{The Tibet hybrid experiment}

 $\;\;\;\;$The Tibet hybrid experiment consists of air-shower array
 (Tibet-III) and YAC-I. The Tibet-III array [1] consists of 733
scintillation detectors (0.5 $m^{2}$ each). Fast-timing detectors
are placed with 7.5 $m$ spacing and density detectors are placed
with 15 $m$ spacing. An event trigger signal is issued when any
four-fold coincidence occurs in FT counters with each of them
recording more than 0.6 particles. The primary energy of each AS
event is determined by the air shower size ($N_e$) which is
calculated by fitting the lateral particle density distribution to
the modified Nishimura-Kamata-Greisen (NKG) structure function.

$\;\;\;\;$YAC-I that consists of 16 EAS core detectors is shown in
Fig.1 and for the brief description see [2],which has started data
taking since May, 2009. YAC-I is located near the center of the
Tibet-III air-shower array, operating simultaneously with Tibet-III.
For the coincident events Tibet-III provides the total energy and
the direction of air showers and YAC-I observes high energy
electromagnetic particles in the core region.

\begin{figure}[!t]
  \vspace{5mm}
  \centering
  \includegraphics[width=3.in]{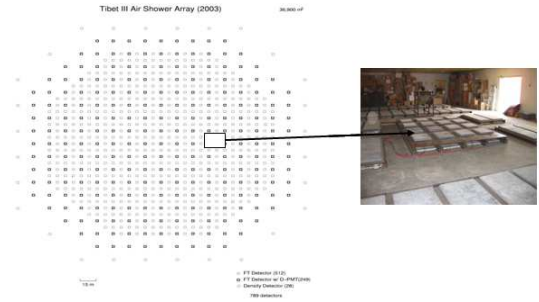}
  \caption{Schematic view of the Tibet-III air-shower array and YAC-I array.}
  \label{simp_fig}
 \end{figure}

$\;\;\;\;$If any one of YAC-I detectors makes a trigger signal that
corresponds to at least 20 MIPs' incidence, all ADC data from all
YAC-I units are recorded. Also the trigger signal is sent to DAQ
system for AS array. ADC modules of YAC-I are calibrated every 4
hours. ADC pedestal values are measured every 10 minutes. Each DAQ
system has GPS clock module independently. The matching between YAC
data and AS data is made using coincidence of GPS clocks and trigger
tag to AS array. The coincidence condition of GPS is about 1 $\mu s$
[3].

$\;\;\;\;$In the following analysis, we present our results based on
the YAC-I data and AS data.

\section{Simulation and Analysis}
$\;\;\;\;$A Monte Carlo simulation has been carried out on the
development of EAS in the atmosphere and the response in YAC-I. The
simulation code CORSIKA (version 6.204)[4] including QGSJET2 and
SIBYLL2.1 hadronic interaction models are used to generate AS
events. The assumed primary cosmic-ray composition in MC is based on
Non-Linear Acceleration (NLA) model (about details, please see
[5][2]). The factional contents of the assumed primary cosmic-ray
flux are listed in Table 1. Primaries isotropically incident at the
top of the atmosphere within the zenith angles from 0 to 60 degrees
are injected into the atmosphere. The minimum primary energy of this
simulation is set at 1 TeV. Secondary particles are traced to the
altitude of 4300 $m$ till 300 MeV. For each  simulated AS event that
reaches the observational level, its core is dropped randomly onto
an area of 52.84 $m$ $\times$ 52.14 $m$, which includes the marginal
space of 25 $m$ outside the each side of detectors. MC simulation
shows that the core resolution is better than 2 $m$ if taking the
$N_b$ weighted center as the AS core. The electromagnetic showers in
the lead layer induced by electrons or photons that hit any detector
unit of the array are treated by a subroutine that is based on the
detector simulation code EPICS (version 8.64)[6].

$\;\;\;\;$Normally, the following quantities of YAC are used to
characterize an EAS core event: The number of shower particles
hitting a detector unit is called 'burst size' ($N_b$). When the
burse size of a detector unit is higher than 200, this unit is
defined as a 'fired' one. We also call the total burst size of all
fired detector units as $\sum{N_b}$, the maximum burst size among
fired detectors as $N_b^{top}$.

\begin{table}[!t]
\begin{center}
\caption{\small{Fractions of components in the assumed primary
cosmic-ray spectrum of the NLA model.}}
\begin{tabular}{l|ccc}
\hline
Com. & $10^{14}$-$10^{15}$ eV &$10^{15}$-$10^{16}$ eV & $10^{16}$-$10^{17}$ eV\\
\hline
\small{P} & 26.3\% & 10.0\% & 5.0\% \\
\small{He} &28.7 \% & 17.5\% & 11.4\% \\
\small{M} & 34.4\% & 50.3\% & 48.5\% \\
\small{Fe} & 10.6\% & 22.2\% & 35.1\% \\ \hline
\end{tabular}
\label{table_single}
\end{center}
\end{table}

\begin{table}[!t]
\begin{center}
\caption{\small{The fraction of the components after the event
selection.}}\label{table_single}
\begin{tabular}{l|cc}
\hline
Com. & $10^5<N_e<5\times10^5$ & $N_e\geq5\times10^5$  \\
\hline

\small{P}  & 58.3\% & 21.3\%  \\
\small{He}  &29.8\% & 23.4\% \\
\small{M}  & 11.3\% & 40.8\%  \\
\small{Fe}  & 0.6\% & 14.5\% \\ \hline
\end{tabular}
\end{center}
\end{table}

\begin{figure}[!t]
  \vspace{3mm}
  \centering
  \includegraphics[width=2.5in]{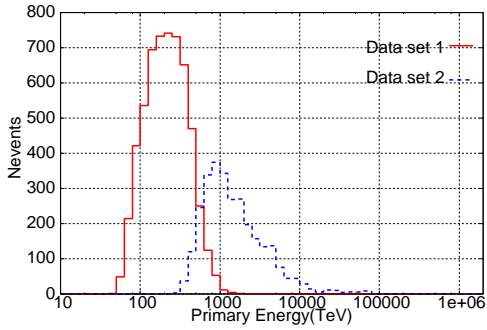}
  \caption{The distribution of primary energy of the sample of two data sets.}
  \label{simp_fig}
 \end{figure}

\begin{table}[!t]
\begin{center}
\caption{\small{The number of events of 2 selected
samples.}}\label{table_single}
\begin{tabular}{l|ccc}
\hline
  Ne  & \small{QGSJET} & \small{SIBYLL}  & \small{Expt.data} \\
\hline
$10^5<N_e<5\times10^5$   & 5687      & 4581     & 523  \\
\hline
$N_e\geq5\times10^5$& 3858     & 2888       & 317 \\
\hline
\end{tabular}
\end{center}
\end{table}

$\;\;\;$We can obtain different event samples that have different
average primary energy and different sample size by using different
threshold of $N_e$. Therefore, we can see how some physics
quantities change with energy simultaneously. We obtain two data
sets by
imposing the following conditions:\\
(1) $N_{b}\geq200$, $N_{hit}\geq6$, $N_b^{top}$$\geq$1500,
 $10^5<N_e<5\times10^5$;\\
(2) $N_b\geq200$,  $N_{hit}\geq6$, $N_b^{top}$$\geq$1500,
 $N_e\geq5\times10^5$;\\
$\;\;\;\;$Fig.2 shows the primary-energy distribution of these two
data sets. The mode energy as known from the Monte Carlo is 260 TeV
and 1800 TeV, respectively.\\
$\;\;\;\;$We sampled $1.8\times10^{10}$ and $1.17\times10^{10}$
primaries for the QGSJET2 and SIBYLL2.1 model, respectively. The
 number of events of two data sets selected under the above
conditions can be seen from Table 3. The simulated data were
analyzed in the same manner as in the procedure for the experimental
data analysis. In present paper, we used the experimental data set
obtained from May, 2009 through January, 2010. An event coincidence
between AS events and YAC-I events is made by their arrival time.
Deadtime correction of 12\% for AS trigger system and 15\% for YAC-I
trigger system are taking into account. The data sample coming from
successful coincidence corresponds a live time of 100.5 days. The
total number of events which is selected under the above two
conditions from the experimental data is also listed in Table 3.

\section{Results and Discussion}

\begin{figure}[!t]
  \vspace{5mm}
  \centering
  \includegraphics[width=2.5in]{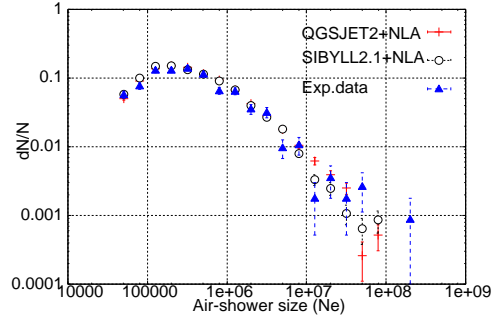}
  \caption{The comparison of air-shower size $N_e$ between MC and
experimental data.}
  \label{simp_fig}
 \end{figure}

 \begin{figure}[!t]
  \vspace{3mm}
  \centering
  \includegraphics[width=2.5in]{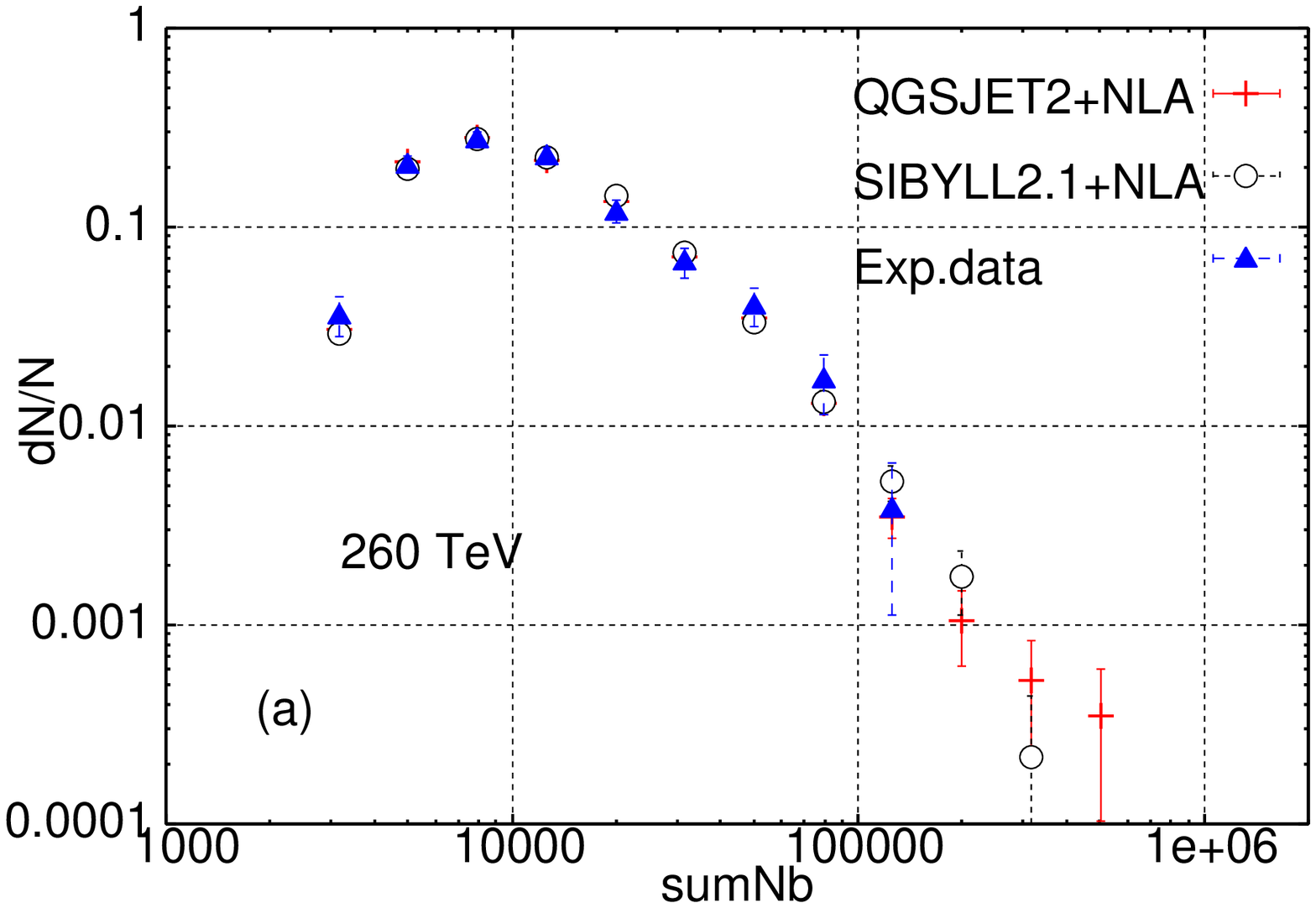}
  \includegraphics[width=2.5in]{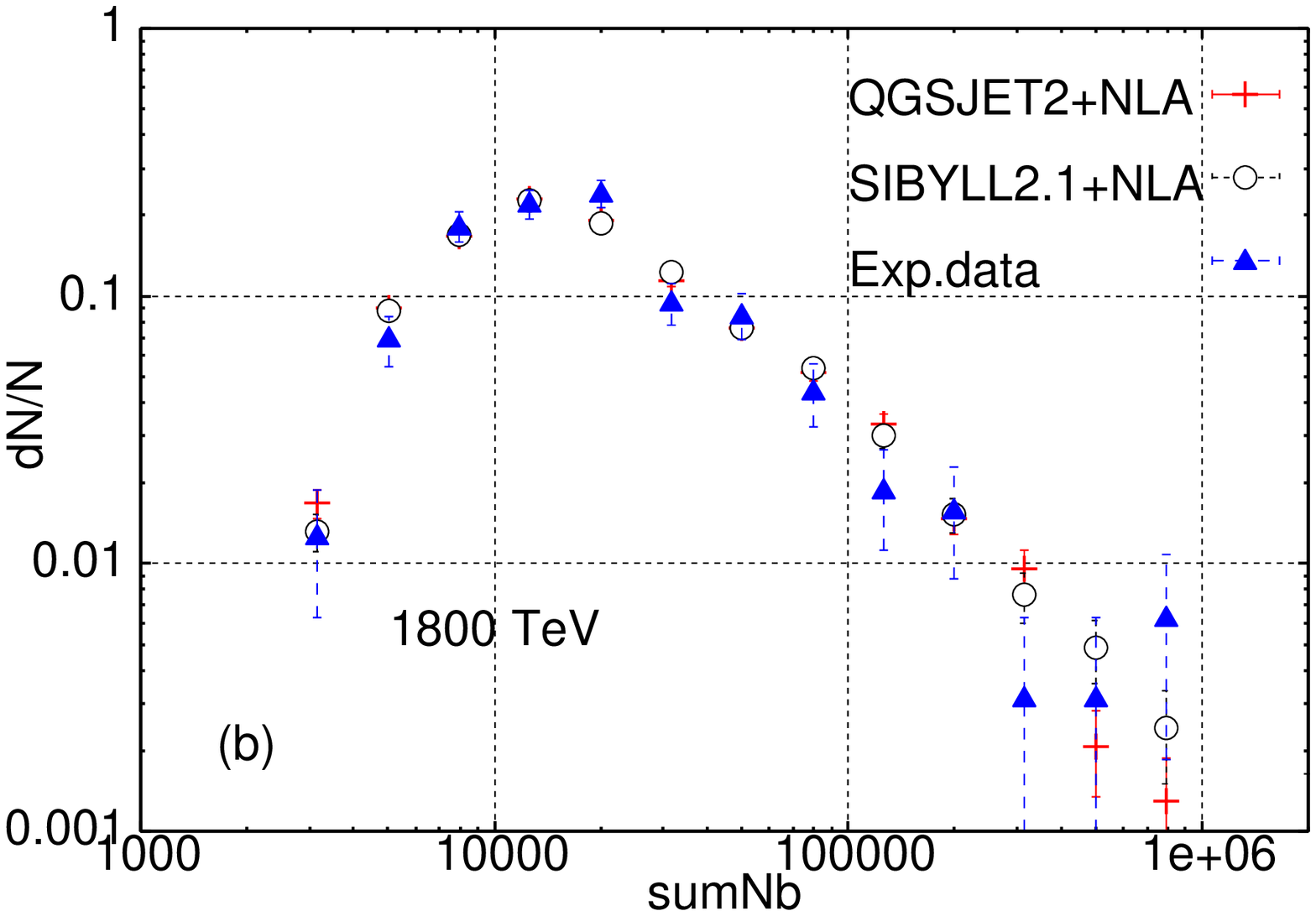}
  \caption{The spectrum of the total burst size $\sum{N_b}$ obatined by MC
   and experimental data  at 260 TeV (a) and 1800 TeV (b) energy region,
   respectively.}
  \label{simp_fig}
 \end{figure}

\begin{figure}[!t]
  \vspace{3mm}
  \centering
  \includegraphics[width=2.5in]{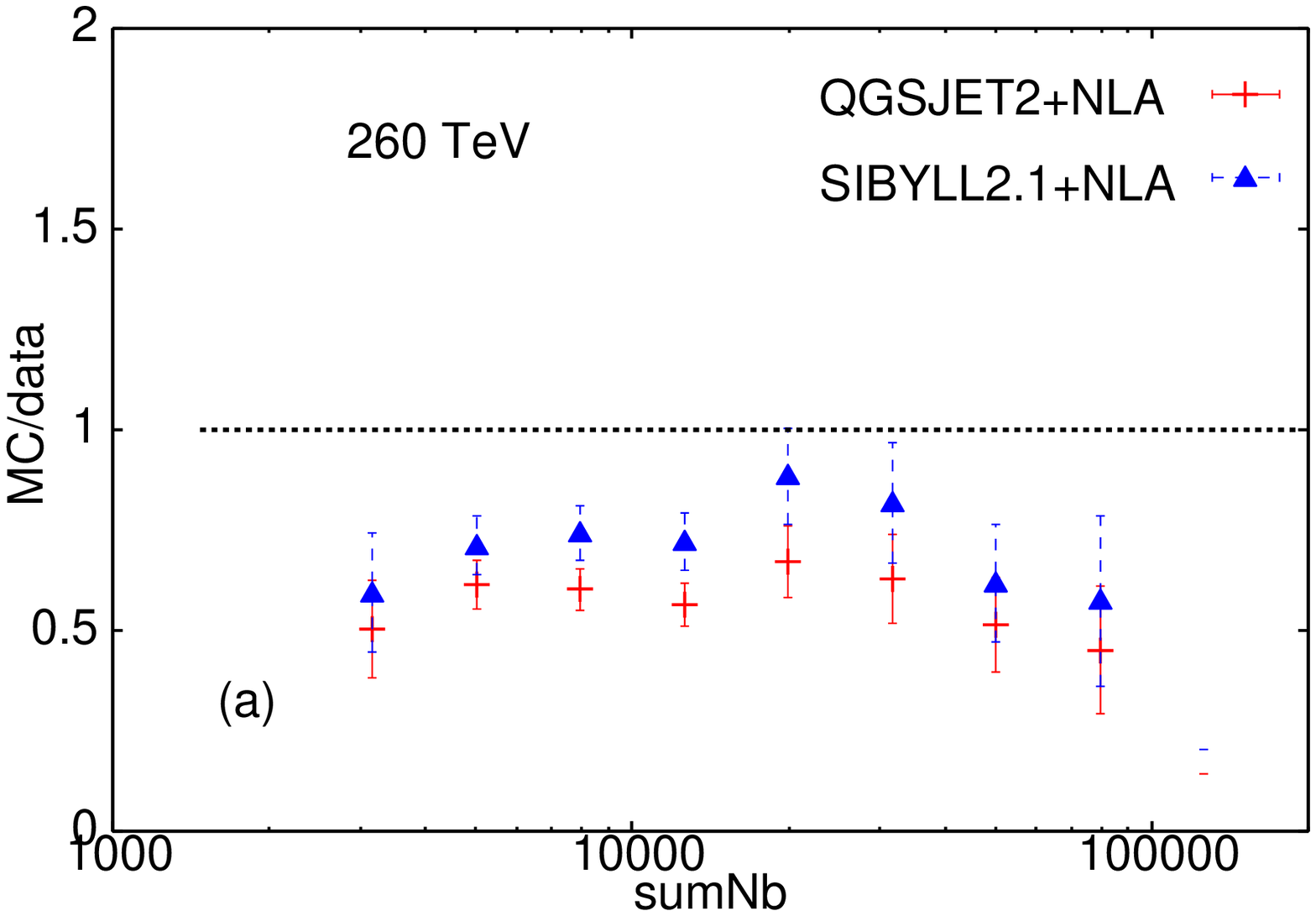}
  \includegraphics[width=2.5in]{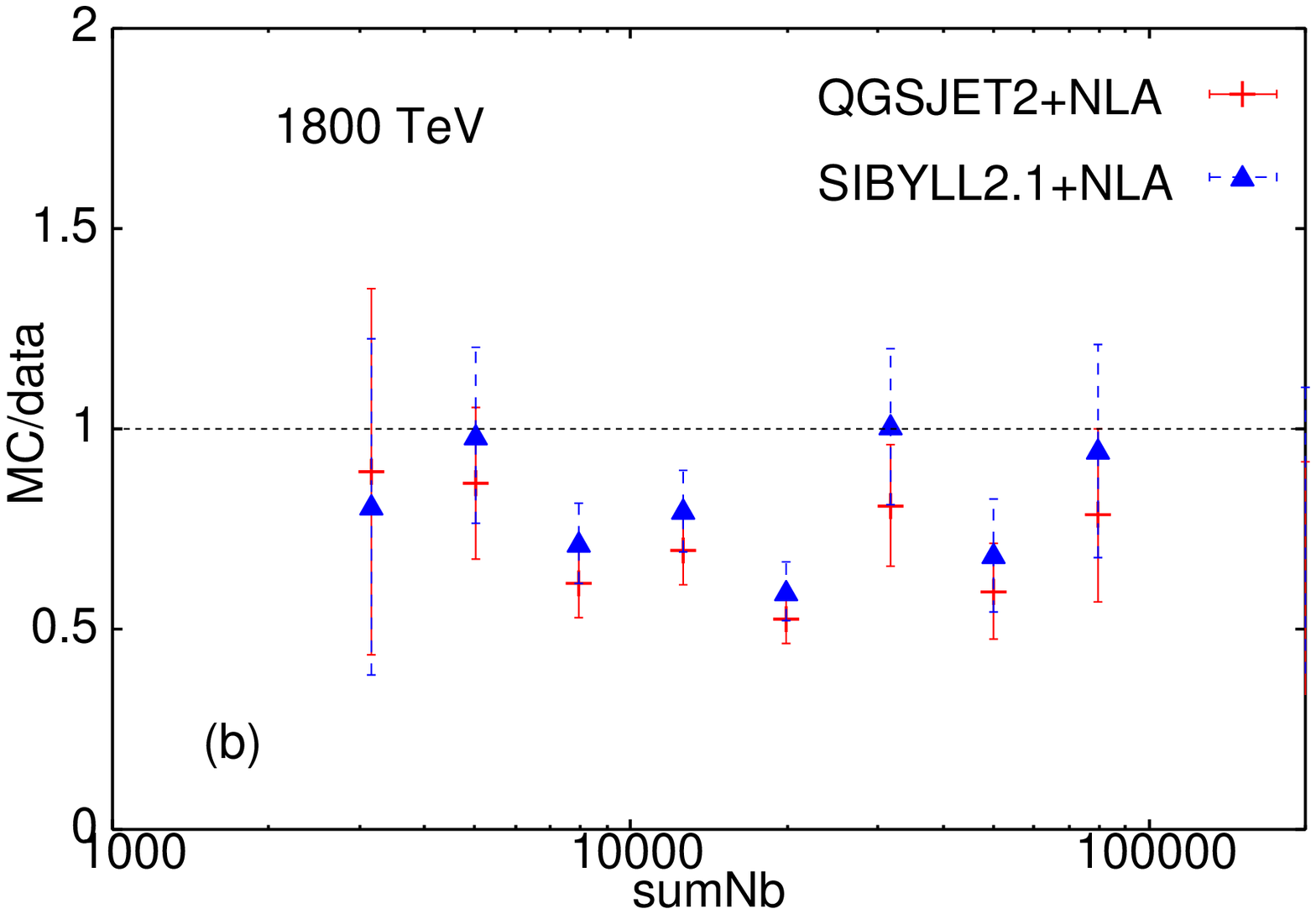}
  \caption{The flux ratio of the absolute intensities of the
  total burst size $\sum{N_b}$ obatined by MC
   and experimental data  at 260 TeV (a) and 1800 TeV (b) energy region,
   respectively.}
  \label{simp_fig}
 \end{figure}

$\;\;\;\;$Since our Monte Carlo simulation is started from 1 TeV, in
order to normalize MC data and experimental data, we need to know
the integral intensity of all particles of cosmic rays at
E$_{0}$$\geq$ 1 TeV: starting from H$\ddot{o}$randal's spectra of
each composition[7], we improve the major 8 ones (P, He, C, O, Ne,
Mg, Si, and Fe) by the newest measurements [8][9][10]. The resultant
integral intensity: I($\geq$1 TeV) = 0.139
$cm$$^{-2}$$s$$^{-1}$$sr$$^{-1}$ with the error +0.0013, -0.0012
coming from the error of the index of each the 8 spectra.\\

$\;\;\;\;$Fig.3 is the comparison of air-shower size ($N_e$) between
MC and experimental data which are normalized by number of events.
It shows both MC models produce air-shower size distribution
consistent with experimental data.\\

$\;\;\;\;$Fig.4 is the comparison of the total burst size
$\sum{N_b}$ which are normalized by number of events between MC and
experimental data at 260 TeV and 1800 TeV energy region,
respectively. $\sum{N_b}$ should depend sensitively on the inelastic
interaction cross section, the inelasticity, and particles produced
in the forward region. It shows that these two hadronic interaction
models have the same shape with experimental results.\\

$\;\;\;\;$Fig.5 is the flux ratio of the absolute intensities
between MC and experimental data. It shows that both QGSJET2 and
SIBYLL2.1 give about 40\% lower flux.

\section{Summary}

$\;\;\;\;$The shape of the distributions of $\sum{N_b}$ is
consistent between the YAC-I data and simulation data in these two
cases, indicating that from 260 TeV to 1800 TeV, the particle
production spectrum of QGSJET2 and SIBYLL2.1 may correctly reflect
the reality within our experimental systematic uncertainty of a level about 10\%.\\

$\;\;\;\;$But note that, NLA composition model used a steeper He
spectrum [5], our results are still affected by the composition
model used, comparing with the new results from PAMELA and CREAM.
Enhancing He spectrum may change the results. The bending energy of
p and He spectra may also be an important factor. It is also noticed
that, seen from Table 2, the 2 data samples have different
composition. Therefore, at present stage, it is not simple to make a
conclusion. A further study is needed and is going on. \\

$\;\;\;\;$The above results show that taking the priority of high
altitude (like Yangbajing) an EAS core event sample can be obtained
with high statistics by using YAC type detector, and the hadronic
interaction models can be checked. YAC-II has been constructed and
start data taking since August 1st, 2011, the more results will be
expected.
\section{Acknowledgements}
This work is supported by the Chinese Academy of Sciences
(H9291450S3) and  the  Key  Laboratory of  Particle  Astrophysics,
Institute of High Energy Physics, CAS.  The Knowledge Innovation
Fund (H95451D0U2 and H8515530U1) of IHEP, China and the project
Y0293900TF of NSFC also provide support to this study.


\clearpage

\end{document}